\documentstyle[12pt]{article}

\newcommand{\ignore}[1]{}

\begin{document}
\sloppy
\sloppy
\sloppy
\vskip 0.5 truecm
{\baselineskip=14pt
 \rightline{
 \vbox{
       \hbox{UT-782}
       \hbox{July  1997}
       }}}
~~\vskip 5mm
 \begin{center}
{\Large{\bf 
An extended q-deformed $su(2)$ algebra 
and \\ the Bloch electron problem}}
\end{center} 

\vskip .2 truecm
\centerline{\bf  Kazuo Fujikawa    
  and  Harunobu Kubo\footnote[1]{ 
 E-mail address : kubo@hep-th.phys.s.u-tokyo.ac.jp 
 } }
\vskip .2 truecm
\centerline {\it Department of Physics,University of Tokyo}
\centerline {\it Bunkyo-ku,Tokyo 113,Japan}
\vskip 0.5 truecm

%%%%%%%%%%%%%%%%%%%%%%%%%%%%%%%%%%%%%%%%%%%%%%%%%%%%%%%%%%%%%%%%%%%%%%%%%%%%

%%%
\begin{abstract}
It is shown that an extended  q-deformed $su(2)$ algebra  with  an extra
(``Schwinger '') term  can describe  Bloch electrons in a  uniform magnetic
field with an 
additional periodic potential.
This is a generalization of the analysis of 
 Bloch electrons by Wiegmann and Zabrodin.
By using a representation theory of this q-deformed algebra,  
we obtain  functional Bethe ansatz equations whose solutions should be
 functions of finite degree. It is also shown that the zero energy solution
 is expressed in terms of an Askey Wilson polynomial.
\end{abstract}
%\large
The quantum deformation \cite{SDJ} of algebras was 
introduced  in connection with inverse scattering problems and 
integrable models. It is also known that this notion  has a  deep relation 
to Yang-Baxter equations \cite{FKS}.
The $U_q(sl_2)$ algebra, one of the q-deformed algebras, was 
applied to an analysis of the two-dimensional Bloch electron problem
\cite{H}
 by Wiegmann and Zabrodin \cite{WZ}. 
They  found  functional Bethe ansatz equations for this problem.   
From the  representation theory of  $U_q(sl_2)$ algebra 
on the functional space, they deduce that 
the solutions of the  Bethe ansatz equations 
should be  functions of  finite degree. 
The zero energy  solutions and the generalized 
 functional Bethe ansatz equations  
have  also been discussed in Refs. \cite{HKW}\cite{FaKa}. 
Biedenharn \cite{B} and 
Macfarlane \cite{M} constructed $U_q(sl_2)$ algebra, in the manner of
Schwinger's construction of conventional $su(2)$,  by using  two sets of
q-deformed  oscillator algebras.
The  q-deformed oscillator  algebra ${\cal H}_q(1)$, if suitably defined, 
supports the  Hopf structure \cite{Y} \cite{OS1}.
A  method to construct a representation of  ${\cal H}_q(1)$, which is 
manifestly free of negative norm, was proposed \cite{FKwO}.
This representation of ${\cal H}_q(1)$ was used to
analyze the phase operator problem \cite{CGN} \cite{E} of the photon by 
using a notion of index \cite{F}.

Recently,  an  extended q-deformed  $su(2)$ algebra was proposed \cite{FKO}
by 
using a representation of the above  q-deformed oscillator algebra, which
is 
manifestly free of negative norm.
This construction is a one parameter generalization
(the Casimir degree of freedom ``$n_0$'' ) of the construction of 
Biedenharn \cite{B} and  Macfarlane \cite{M}.
It is reduced to conventional $U_q(sl_2)$  if one  fixes the  Casimir to a
specific value. 
This algebra has an additional term in commutation relations of 
$U_q(sl_2)$,
which we tentatively called as `` Schwinger term'', since  this extra term
was 
originally introduced to maintain the 
representation
 free of negative norm for the  deformation parameter  $q$ at  a  root 
of unity \cite{FKO}.
This extended q-deformed $su(2)$ algebra 
  can be realized on a  functional space 
which is analogous to the conventional functional 
realization of $U_q (sl_2)$, but it contains an  additional one parameter 
degree of freedom corresponding to the Casimir freedom .  
In a manner similar to the case of  $U_q(sl_2)$ algebra, 
we can construct a cyclic representation from 
the functional realization of the extended q-deformed $su(2)$ algebra. 
The non-linear correspondence\cite{K} between this algebra 
and  q-deformed oscillator algebra 
was   found in their cyclic representations 
and functional realizations \cite{P} \cite{ACDI}. 
Other aspects of q-deformed oscillator algebra and the extended q-deformed
$su(2)$
algebra have also been discussed  in Ref \cite{OS3}.

In this paper, we show that the extended q-deformed $su(2)$ algebra is used
to describe  a one-parameter generalization of the 
Wiegmann and Zabrodin  analysis  of  the Bloch electron problem,
in which  an extra periodic potential appears in addition to the uniform 
magnetic field.
Firstly,  we summarize the notations of  the extended q-deformed $su(2)$ 
algebra  with a ``Schwinger term''.
Defining relations of the algebra are, 
\begin{eqnarray}\label{def}
{[}S_{\pm}, S_{3}{]} &=& \mp S_{\pm},\nonumber\\
{[}S_{+}, S_{-}{]}&=& [2S_{3}] + 4[n_{0}]\sin \pi\theta \sin 2\pi\theta 
[S_{3}][{\cal C} + \frac{1}{2}],
\end{eqnarray}
where 
\begin{equation}
[x]=\frac{q^x-q^{-x}}{q-q^{-1}}.
\end{equation}
The operator ${\cal C}$ commutes with all other quantities, and  ${\cal C}$
is fixed at 
$2{\cal C} +1 = 2j+1-2n_{0}$ for a $2j+1$ dimensional
representation
of this algebra \cite{FKO}. 
We note that the algebra (1) for 
\begin{equation}
q = e^{2\pi i\theta},
\end{equation}
with an arbitrary real $\theta$ and for an arbitrary number $n_{0}$ has a
$2j + 1$ dimensional highest weight representation. 

The $2j+1$ dimensional highest 
weight representation of 
the algebra (\ref{def}) can also be realized by $q-$difference equations as
\begin{eqnarray}\label{diff}
\tilde{S}_{+}\psi(z)&=&(q-q^{-1})^{-1}z(q^{2j-n_{0}}\psi(q^{-1}z)-
q^{-2j+n_{0}}\psi(qz))+z[n_0]\psi(z), \nonumber \\
\tilde{S}_{-}\psi(z)&=&-(q-q^{-1})^{-1}z^{-1}(q^{n_{0}}\psi(q^{-1}z)-
q^{-n_{0}}\psi(qz)) +z^{-1}[n_0]\psi(z),  \\
q^{\tilde{S}_{3}}\psi(z)&=&q^{-j}\psi(qz), \nonumber 
\end{eqnarray}
where $\psi(z)$ is a polynomial of degree $2j$.
This representation satisfies 
the highest weight condition $\tilde{S}_{+}z^{2j}=0$ and the 
lowest weight condition $\tilde{S}_{-}\cdot 1 =0$. 
The representation  of (\ref{diff}) for the bases, 
$z^{j+m},\ m= j, j-1, ......, -j$, is given by
\begin{eqnarray}
\tilde{S}_{+}z^{j+m} &=& ([j - m - n_{0}] + [n_{0}])z^{j+m+1},\nonumber\\
\tilde{S}_{-}z^{j+m} &=& ([j + m - n_{0}] + [n_{0}])z^{j+m-1},\nonumber\\
q^{\tilde{S}_{3}}z^{j+m} &=& q^{m}z^{j+m}.
\end{eqnarray}
This representation is related to the standard  $2j +1$ dimensional 
representation $ S_{\pm}, S_{3}$ by a non-unitary similarity transformation
$ S_{\pm}= A^{-1}\tilde{S}_{\pm}A ,  S_{3}= \tilde{S}_{3}$ \cite{FKO}.
We will come back to this point later. 
For a specific deformation parameter $q = \exp [\pi i P/Q]$ for mutually 
co-prime integers $P$ and $Q$ ,  we can define a
cyclic representation, which is   obtained by putting  
$z=q^{k}, (k=1,2,\cdots,2Q)$, in (\ref{diff}).
There are $2Q$ bases $\psi_{k}\equiv \psi(q^{k})$ which satisfy 
$\psi_{k+2Q}=\psi_{k}$, and we have 
\begin{eqnarray}\label{cyclic}
\rho (S_{+})\psi_{k}&=& (q-q^{-1})^{-1}(- q^{k-2j+n_{0}}\psi_{k+1}
                                     + q^{k+2j-n_{0}}\psi_{k-1})
                 +q^{k}[n_0]\psi_{k},  \nonumber\\
\rho (S_{-})\psi_{k}&=&(q-q^{-1})^{-1}(q^{-k-n_{0}}\psi_{k+1}
                                     -q^{-k+n_{0}}\psi_{k-1})
                 +q^{-k}[n_0]\psi_{k}, \nonumber\\
q^{\rho (S_{3})}\psi_{k}&=&q^{-j}\psi_{k+1}. 
\end{eqnarray}
It is confirmed 
that this cyclic representation $\rho(S)$ satisfies the algebra
(\ref{def}). Note that  $2j+1$, $P$ and $Q$ are independent integers. 
In the original construction in \cite{FKO}, we chose $n_{0}$ such that 
\begin{equation}\label{po}
[n_{0}] = \frac{\sin \pi n_{0} \frac{P}{Q}}{\sin \pi \frac{P}{Q}}
        = \frac{1}{|\sin \pi \frac{P}{Q}|},
\end{equation}        
to avoid the negative norm in
the standard $2j+1$ dimensional representation. 

We now consider  tight-binding two-dimensional electrons in a magnetic
field with an
additional periodic  potential
\begin{eqnarray}\label{Ham}
H&=& V_1 T_x +V_1 T_x^{\dagger} +V_2 T_y +V_2 T_{y}^{\dagger}
     +V_3 U+V_3^{*}U^{\dagger},
\end{eqnarray}
where the parameters  $V_1$ and  $V_2$ are real, and  $V_3$ is a complex 
number.
\begin{equation}
T_x= \sum_{m,n} c^{\dagger}_{m+1,n}c_{m,n} e^{i \theta^{x}_{m,n}},\quad
T_y= \sum_{m,n} c^{\dagger}_{m,n+1}c_{m,n} e^{i \theta^{y}_{m,n}}, \quad
 U = \sum_{m,n}q^{ m+n }c^{\dagger}_{m,n}c_{m,n}.
\end{equation}
Here  $c_{m,n}$ is the annihilation operator for an electron at site
$(m,n)$,
and we choose  a very specific diagonal potential $U$ containing the 
parameter $q$ for later convenience. We assumed that the constant phase
factors 
of  $V_1$ and  $V_2$ can be absorbed into the gauge potentials 
$\theta^{x}_{m,n}$ and $\theta^{y}_{m,n}$.
The gauge potentials  $\theta^{x}_{m,n}$ and $\theta^{y}_{m,n}$ are related
to
a flux per plaquette $ \phi_{m,n}$ at $(m,n)$ by
\begin{equation}
rot_{m,n}\theta = \Delta_x \theta^{y}_{m,n} - \Delta_y\theta^{x}_{m,n}
= 2\pi \phi_{m,n},
\end{equation}
where the difference operators  $\Delta_x$ and  $\Delta_y$ operate on a
lattice function $f_{m,n}$ as
\begin{equation}
 \Delta_x f_{m,n} =  f_{m+1,n}-  f_{m,n}, \quad
 \Delta_y f_{m,n} =  f_{m,n+1}-  f_{m,n}.
\end{equation}

The  Hamiltonian (\ref{Ham}) is invariant under the following gauge
transformation,
\begin{equation}
c_i \to \Omega_i c_i, \quad
e^{i \theta_{i,j}} \to  \Omega_i e^{i \theta_{i,j}} \Omega_j^{-1}, \quad
|\Omega_i | =1. 
\end{equation}
In the following, we  assume that the magnetic field is uniform and
rational,
$-\phi_{m,n}=\phi =P/Q$ with mutually prime integers $P$ and $Q$.
We can then take a diagonal gauge specified by
\begin{equation}\label{diag}
\theta^{x}_{m,n}= \pi \phi ( m+n), \quad 
\theta^{y}_{m,n}= -\pi \phi ( m+n+1).
\end{equation}
In this diagonal gauge, if we choose the  parameter $q$, which is later
identified 
with the deformation parameter of q-deformed algebra,  as a root
of unity 
\begin{equation}
q^{2Q}=1, \quad  q=e^{i\pi \frac{P}{Q}}, \quad P,Q \in Z,  
\end{equation}
the Hamiltonian becomes  periodic along $(1,1)$ and $(1,-1)$ directions;
the period in $(1,1)$ direction is $2Q$.
We can then use the 
Bloch theorem with momenta  $p_+$ and $p_-$ for {\em one particle} states 
\begin{equation}
| \Phi (p_+, p_-) \rangle = \sum_{m,n}\Psi_{m,n}(p_+, p_-)c^{\dagger}_{m,n}
| 0 \rangle,
\end{equation}  
where 
\begin{eqnarray}
&&  \Psi_{m,n}({\bf p} )= e^{ip_- ( m-n) + ip_+ ( m+n)} 
\psi_{m+n}({\bf p }),
\\
&&\psi_{k+ 2Q}({\bf p })=\psi_{k}({\bf p }), \quad k=1, \cdots , 2Q. 
\end{eqnarray}
Note that the "periodic potentials"  in the present problem (8) depend only
on the 
combination $m+n$ which is the reason why we have $\psi_{m+n}({\bf p})$. 
The Schr\"{o}dinger equation, 
$H | \Phi ({ \bf p}) \rangle   =E | \Phi ({ \bf p}) \rangle $,  is written
on the basis $\psi_{k}$ as follows
\begin{eqnarray}\label{Sch}
  (V_1 q^{k-1} e^{-i(p_- + p_+)}+V_2 q^{-k}  e^{i(p_- - p_+ )})\psi_{k-1} 
+(V_1 q^{-k} e^{i(p_- + p_+)} +V_2 q^{k+1} e^{-i(p_- - p_+)})\psi_{k+1}
&&    \nonumber \\
+(V_3 q^{k} +V_3^{*}q^{-k})\psi_{k}
   = E \psi_{k}, 
&&
\end{eqnarray}
where, by noting  $\phi = P/Q$,  we used a relation,
\begin{equation}
q=e^{i \pi \frac{P}{Q}} =  e^{i\pi \phi}.
\end{equation}
Inserting $z=q^{k}$ into  the following functional equation
and identifying  $\psi_{k} =  \psi (q^{k})$,
\begin{eqnarray}
  (V_1 zq^{-1} e^{-i(p_- + p_+)}+V_2 z^{-1}  e^{i(p_- - p_+
)})\psi(q^{-1}z) 
    +(V_1 z^{-1}e^{i(p_- + p_+)} +V_2 z q e^{-i(p_- - p_+)})\psi(qz) 
&&    \nonumber \\
   +(V_3 z +V_3^{*}z^{-1})\psi(z)
   = E \psi(z), 
&&
\end{eqnarray}
 we  recover the  original Schr\"{o}dinger equation (\ref{Sch}). In
connection 
 with our analysis below,  
it is crucial that  a generic
 $\psi(z)$ is not always written in the form of an  integral power in $z$
of finite degree. 
In order to get such a  special   $\psi(z)$ of a finite polynomial  in $z$,
the 
parameters of the model need to be  restricted.

%%%%%%%%%%

Next we show that the Hamiltonian (\ref{Ham}) with  specific values of
parameters can
be
described in terms  of the generators of  the extended  
$su_q(2)$.  
This is one of the cases in which 
the functional space consists of finite polynomials in $z$.
In the diagonal gauge (\ref{diag}), we can write the Hamiltonian 
as a  linear combination of generators of  the extended $su_q(2)$ .

We  postulate the  following form of Hamiltonian with undetermined
parameters $a$ and $\theta_{\alpha}$,
\begin{equation}\label{HamS}
H=i(q-q^{-1}) a( e^{-i\theta_{\alpha}} S_-(n_0) + e^{i\theta_{\alpha}}
S_+(n_0)),
\end{equation}
with positive $a > 0$  and  real $\theta_{\alpha}$ to maintain $H$
hermitian
in the standard $2j+1$ dimensional representation. Our  $H$ is a
generalization of the  Hamiltonian in 
Ref.\cite{WZ} for $U_{q}(sl_{2})$.  Note that we always have $S_-(n_0) =
S_+(n_0)^{\dagger}$ in the standard $2j+1$ dimensional representation 
if (7) 
is satisfied \cite{FKO}, whereas $S_- = S_+^{\dagger}$ is not ensured in
general for 
$U_{q}(sl_{2})$ with $n_{0} =0$. 

On the functional space the Hamiltonian (\ref{HamS}) acts as 
\begin{equation}
H\Psi(z)=i(q-q^{-1})a( e^{-i\theta_{\alpha}} S_-(n_0) +
e^{i\theta_{\alpha}} S_+(n_0))
\Psi(z)=E\Psi(z).
\end{equation} 
The explicit form of this functional equation  is written as (see eq.(4))
\begin{eqnarray}\label{HamBA}
&& ia(- e^{i\theta_{\alpha}}q^{1+n_0}z + 
 e^{-i\theta_{\alpha}}q^{-n_0}z^{-1})\Psi(qz)
   +ia( e^{i\theta_{\alpha}}q^{-1  -n_0}z - e^{-i\theta_{\alpha}}
q^{n_0}z^{-1})\Psi(q^{-1}z)
\nonumber \\
&& +ia( e^{-i\theta_{\alpha}}z^{-1} + e^{i\theta_{\alpha}}z)
(q^{n_0}-q^{-n_0})\Psi(z)=E\Psi(z),
\end{eqnarray}
where we assumed  a very specific representation with  $q^{2j+1}=1$,
which is satisfied by
\begin{equation}
2j+1 = Q\ \ and \ \ P=even ,
\end{equation}
or
\begin{equation}
2j +1 = 2Q ,
\end{equation}
for $q=e^{i\pi P/Q}$. Here we utilized the fact that $2j+1$ and $Q$ are
generally independent 
integers. The choice $2j+1 = Q$, which is also the choice in Ref.
\cite{WZ},
suggests an integral $j$ whereas $2j+1=2Q$ suggests half an odd integer
$j$.
If we identify eq.(20) with eq.(23), we obtain 
\begin{eqnarray}
\label{COM1}
&& V_1 e^{-i(p_- + p_+)} = i a e^{i\theta_{\alpha}}q^{-n_0}, 
\nonumber\\
&& V_2 e^{i(p_- - p_+)} =  -i a e^{-i\theta_{\alpha}}q^{n_0}, 
\nonumber\\
&& V_3  = i a e^{i\theta_{\alpha}}(q^{n_0}- q^{-n_0}), 
\nonumber\\
&& V_1 e^{i(p_- + p_+)} = i a e^{-i\theta_{\alpha}}q^{-n_0},
\nonumber\\
&& V_2 e^{-i(p_- - p_+)} = -i a e^{i\theta_{\alpha}}q^{+n_0}, 
\nonumber\\
&& V_3^*  = i a e^{-i\theta_{\alpha}}(q^{n_0}- q^{-n_0}). 
\end{eqnarray}
It can be confirmed that the conditions in (26) are satisfied by the
following 
choice 
of parameters:
\begin{eqnarray}\label{ex1}
&& V_{1} = V_{2}= a, \nonumber\\
&&p_{+}  = \pi m_{1},\nonumber\\
&&p_{-} = -\theta_{\alpha} + \pi m_{2},
\nonumber\\
&& V_{3} \equiv 2a e^{i\theta_{3}} = -2a e^{i\theta _{\alpha}}\sin (\pi
n_{0}
P/Q),
\end{eqnarray}
with suitable $m_{1}, m_{2} \in Z$. In the last relation of these
equations,
we recalled  $\sin \pi n_{0} P/Q  =\pm 1$ or $q^{n_{0}} = \pm i$ due to the
definition in (7).

The extra potential terms in the Hamiltonian 
(\ref{Ham}) are given by 
\begin{eqnarray}
V_3 U + V_3^{*}U^{\dagger}
=4a \sum_{m,n}
\cos \{\pi \frac{P}{Q}(m+n) +\theta_3 \}
c^{\dagger}_{m,n}c_{m,n}. 
\end{eqnarray}
This potential has a system size  period $2Q$, and its periodicity 
is in the  diagonal direction. The origin of the potential is shifted by 
the phase factor coming from $V_3$. ( A way to realize this
potential  physically may be to super-impose an electric field, whose time
variation is
very slow.)

Now we construct Bethe ansatz equations. 
We know that the  representation space of an extended $su_q(2)$ algebra 
is a polynomial of  degree $2j$ , namely,  it can be expressed as
\begin{equation}\label{pol}
\Psi(z) = \prod^{2j}_{m=1} (z-z_m).
\end{equation} 
We come back to the original expression of our equation (\ref{HamBA}).
By noting  (\ref{pol}) we can rewrite this Eq.(\ref{HamBA}) as
\begin{eqnarray}\label{Diff}
&&  ia(-e^{i\theta_{\alpha}}q^{1+n_0}z
+e^{-i\theta_{\alpha}}q^{-n_0}z^{-1})
       \prod^{2j}_{m=1}\frac{qz-z_m}{z-z_m}
\nonumber \\
&&  +ia(e^{i\theta_{\alpha}} q^{-1-n_0}z - e^{-i\theta_{\alpha}}
q^{n_0}z^{-1})
       \prod^{2j}_{m=1}\frac{q^{-1}z-z_m}{z-z_m}
\nonumber \\
&& +ia( e^{-i\theta_{\alpha}} z^{-1}+ e^{i\theta_{\alpha}} z) 
(q^{n_0}-q^{-n_0})
   =E.
\end{eqnarray}
Let us assume that our representation is "irreducible" in the sense that 
all the $2j+1$ terms in polynomials of z appear in $\Psi (z)$ in (29),
which 
amounts to all $z_{m} \neq 0$.
The right-hand side  of  Eq.(30) is a constant and has no pole in
$z$, then the left-hand side  
should also be free of  poles in $z$. 
By this  pole free condition,
we obtain Bethe ansatz equations 
\begin{equation}\label{BA}
\frac{e^{2i\theta_{\alpha}}z^{2}_{l} - q^{2n_0 +1  }}
{q^{2n_0 +1}e^{2i\theta_{\alpha}} z^{2}_{l} +1}
=-\prod^{2j}_{m=1}
\frac{qz_{l} - z_m}{z_{l}-qz_m }, \qquad  l=1,
\cdots , 2j.
\end{equation}
The energy eigenvalue  is then given by 
\begin{equation}
E=-ia e^{i\theta_{\alpha}}
(q^{n_0}-q^{-n_0} -q^{n_0-1} + q^{-n_0+1})\sum^{2j}_{m=1}z_m,
\end{equation}
by using the solutions of (\ref{BA}).

Finally,  we discuss  the possibility of writing the zero energy  solution
of  Eq.(\ref{HamBA}) in terms of an  Askey Wilson polynomial\cite{G}.
 The Askey Wilson polynomial $P_m(w)$ , which is a polynomial of degree $m$
in 
$(w + w^{-1})$, is defined by  
\begin{eqnarray}\label{AW}
&&A(w)P_m(q^{2}w) +A(w^{-1})P_m(q^{-2}w) -(A(w)+ A(w^{-1}))P_m(w) \nonumber
\\
&&=(q^{-2m} -1 ) ( 1-abcdq^{2m-2})P_m(w),
\end{eqnarray}
where
\begin{eqnarray}
A(w)=\frac{(1-aw)(1-bw)(1-cw)(1-dw)}{(1-w^{2})(1-q^{2}w^{2})}.
\end{eqnarray}
Let us set $a=-b=q$, $c=-d$ and  replace $q$  by  $q^{1/2}$ in (\ref{AW}) ;
$c$ may  depend on $q$ and  we replace $c(q)\to c(q^{1/2})\equiv \bar{c}$. 
We then have 
\begin{eqnarray}\label{AW1}
(1-\bar{c}^2 w^2)P_m(qw) +(\bar{c}^2 - w^2)P_m(q^{-1}w) 
-(1-w^2)(q^{-m}+\bar{c}^2 q^m )P_m(w)=0. 
\end{eqnarray}
We introduce the function  $\Psi_m(w)$ which is 
a polynomial of degree $2m$ in $w$ by 
\begin{equation}
\Psi_m(w)=w^m P_m(w),
\end{equation}
and replace $w$ in (\ref{AW1})  by $w=\xi z$. We then obtain 
\begin{eqnarray}\label{AW3}
(1-\bar{c}^2\xi^2 z^2)\Psi_m(q\xi z) +(\bar{c}^2 - \xi^2 z^2)
q^{2m}\Psi_m(q^{-1}\xi z) 
-(1-\xi^2 z^2)(1+\bar{c}^2 q^{2m} )\Psi_m(\xi z)=0. 
\end{eqnarray}
If we equate  (\ref{HamBA}) with $E=0$  and (\ref{AW3}) by identifying 
$\Psi (z) = \Psi_{m}(\xi z)$, we have the conditions
\begin{eqnarray}
\label{AWC1}
&&\bar{c}^2 \xi^2 = e^{2i\theta_{\alpha}}q^{1+2n_0},
\nonumber\\
&&\bar{c}^2 q^{2m} = -q^{2n_0},
\nonumber\\
&& \xi^2 q^{2m} = -e^{2i\theta_{\alpha}}q^{-1},
\nonumber\\
&&1+\bar{c}^2 q^{2m}=-(q^{2n_0} -1), 
\nonumber\\
&&\xi^{2}(1+\bar{c}^2 q^{2m})=e^{2i\theta_{\alpha}}(q^{2n_0} -1), 
\end{eqnarray}
which are solved by 
\begin{equation}
\xi^{2} = - e^{2i\theta_{\alpha}},\ \ \   
\bar{c}^{2} = - q^{1+2n_{0}},\ \ \  
q^{2m} = q^{-1},
\end{equation}
for $q^{2n_{0}} \neq 1$ ( Eq.(7) suggests $q^{2n_{0}} =- 1$).
The last relation in (39) gives $q^{2m+1}=1$, which is consistent with
our choice in (24) if $2m=2j$. Namely we have $\Psi (z) = \Psi_{j}(\xi z)$,
which is in fact a polynomial in $z$ of degree $2j$. We have thus
established 
that the zero energy solution of  (23) for the choice of parameters in (24)
is
expressed by an Askey Wilson polynomial.

 In passing we note that the limit  $n_{0}\rightarrow 0$ is usually
singular, due to the 
condition (7), and therefore we cannot recover the results of Wiegmann and 
Zabrodin by a simple limit $n_{0} \rightarrow 0$ of our final results. If
one 
relaxes the condition (7) , on the other hand, the
negative norm generally
appears in the  level of standard $2j+1$ dimensional representation.
However,
it appears that this does not necessarily imply the  negative norm in the
Bloch
electron problem:  Due to the non-unitary similarity transformation $A$,
the state
vectors in the Bloch electron problem correspond to $\langle b|A^{-1}$ and 
$A|k\rangle$, respectively, in terms of the bra $\langle b|$ and ket
$|k\rangle$
vectors in the standard  $2j+1$ dimensional representation, as was noted in
connection with Eq.(5).

To summarize our consideration, the Hamiltonian of Bloch
electrons in a  uniform magnetic field with an additional  periodic
potential can
be expressed as a linear combination of the generators of the extended
q-deformed
$su(2)$ algebra.  The realization of this
q-deformed
$su(2)$ algebra on the functional space leads to  functional Bethe
ansatz equations.  
The solutions of the Bethe ansatz equations (\ref{BA})
are  functions  of finite degree for  a generic value of $n_0$. 
It has also been shown that the zero energy solution is expressed by an
Askey 
Wilson polynomial, but we have $e^{ip_{+}} = \pm 1 $ instead of the 
mid-band condition $e^{ip_{+}} = i$ in \cite{WZ}.
 
One of the  important aspects of the Bloch electron problem is  
multi-fractal behavior in the  energy spectrum.
In our Hamiltonian (\ref{HamS}) with a generic $n_0$, 
the system could be  at  the off- critical point without  the 
multi-fractal
behavior\cite{T}
\footnote[2]{ We thank Y. Hatsugai for pointing this out.}.
 This property should be checked  by a numerical diagonalization  of the
Schr\"{o}dinger equation (\ref{Sch}). \\ 
{\bf Acknowledgments }\\
 We would like to thank 
Y. Hatsugai, Y. Morita, C. H. Oh, and J. Shiraishi  for  valuable
suggestions and
discussions. 
%%%%%%%%%%%%%%%%%%%%%%%%%%%%%%%%%%%%%%%%%%%%%%%%%%%%%%%%%%%%%%%%%%%%


\begin{thebibliography}{99}
\bibitem{SDJ}
E. K. Sklyanin, Usp. Mat. Nauk {\bf 40}, 214(1985).\\
V. G. Drinfeld, Dokl. Acad. Nauk {\bf 283}, 1060(1985).\\
M. Jimbo, Lett. Math. Phys. {\bf 10}, 63(1985).
\bibitem{FKS}
L. D. Faddeev, {\em Les Houches Lectures} 1982(Elsevier, Amsterdam,
1984).\\
P. P. Kulish and E. K. Sklyanin,\  {\em Lecture Note in Physics} Vol.
151(Springer, Berlin,1982).
\bibitem{H}
D. R. Hofstadter, Phys. Rev.  {\bf  B 14}, 2239(1976).
\bibitem{WZ}
P. B. Wiegmann and A. V. Zabrodin, Phys. Rev. Lett. {\bf 72}, 1890(1994);
Nucl.
Phys. {\bf B422}[FS], 495(1994).
\bibitem{HKW}
Y. Hatsugai, M. Kohmoto, and Y. S. Wu, Phys. Rev. Lett. {\bf 73},
1134(1994); 
Phys. Rev. {\bf B 53}, 9697(1996).
\bibitem{FaKa}
L. D. Faddeev and R. M. Kashaev, Comm. Math. Phys. {\bf 169},181(1995).
\bibitem{B}
L. C. Biedenharn, J. Phys. A: Math. Gen. {\bf 22}, L873(1989). 
\bibitem{M}
A. J. Macfarlane, J. Phys. A: Math. Gen. {\bf 22}, 4581(1989).
\bibitem{Y}
Hong Yan, J. Phys. A:Math. Gen. {\bf 23}, L1150(1990).
\bibitem{OS1}
C. H. Oh and K. Singh, J. Phys A:Math. Gen. {\bf 27}, 5907(1994).     
\bibitem{FKwO}
K. Fujikawa, L. C. Kwek, and C. H. Oh, Mod. Phys. Lett. {\bf A10},
2543(1995).
\bibitem{CGN}
S. H. Chiu, R. W. Gray and C. Nelson, Phys. Lett. {\bf A164}, 237(1992).
\bibitem{E}
D. Ellinas, Phys. Rev. {\bf A45}, 3358(1992).
\bibitem{F}
K. Fujikawa, Phys. Rev. {\bf A52},3299(1995).
%\bibitem{OS2}
%C. H. Oh and K. Singh,  Lett. Math. Phys. {\bf 36}, 77(1996).
\bibitem{FKO}
K. Fujikawa, H. Kubo and C. H. Oh,   Mod. Phys. Lett.  {\bf A 12},
403(1997):
hep-th/9610164.
\bibitem{K}
H. Kubo, Mod. Phys. Lett. {\bf A 12}, 1335(1997):hep-th/9704100.
\bibitem{P}
M. Pillin, Comm. Math. Phys. {\bf 180},23(1996).
\bibitem{ACDI}
A. A. Andrianov, F. Cannata, J.-P. Dedonder and  M. V. Ioffe,
Phys. Lett.  {\bf A 217}, 7(1994).
\bibitem{OS3}
L. C. Kwek and C. H. Oh,  q-alg/9707024.
\bibitem{G}
G. Gasper and M. Rahman, {\em Basic Hypergeometric Series },
(Cambridge Univ. Press, 1990).
\bibitem{T}
D. J. Thouless, Phys. Rev. B. {\bf 28}, 4272(1983).


\end{thebibliography}
\end{document}